\documentclass[twocolumn]{aastex631}
\usepackage{amsmath}
\usepackage[utf8]{inputenc}
\usepackage{cleveref}

\begin{document}



\title{Investigation on Quasi-periodic Oscillation Phase Lag of RE J1034+396}
\correspondingauthor{Wen-Zhong Li}
\email{liwz@ihep.ac.cn}
\correspondingauthor{Shu Zhang}
\email{szhang@ihep.ac.cn}
\correspondingauthor{Qing-Cang Shui}
\email{shuiqc@ihep.ac.cn}

\author[0009-0001-3113-586X]{Wen-Zhong Li}
\affiliation{State Key Laboratory of Particle Astrophysics, Institute of High Energy Physics, Chinese Academy of Sciences, 100049 Beijing, People's Republic of China}
\affiliation{University of Chinese Academy of Sciences, Chinese Academy of Sciences, 100049 Beijing, People's Republic of China}
\author{Shu Zhang}
\affiliation{State Key Laboratory of Particle Astrophysics, Institute of High Energy Physics, Chinese Academy of Sciences, 100049 Beijing, People's Republic of China}
\author[0000-0001-5160-3344]{Qing-Cang Shui}
\affiliation{State Key Laboratory of Particle Astrophysics, Institute of High Energy Physics, Chinese Academy of Sciences, 100049 Beijing, People's Republic of China}
\author{Yu-Peng Chen}
\affiliation{State Key Laboratory of Particle Astrophysics, Institute of High Energy Physics, Chinese Academy of Sciences, 100049 Beijing, People's Republic of China}
\author{Shuang-Nan Zhang}
\affiliation{State Key Laboratory of Particle Astrophysics, Institute of High Energy Physics, Chinese Academy of Sciences, 100049 Beijing, People's Republic of China}
\affiliation{University of Chinese Academy of Sciences, Chinese Academy of Sciences, 100049 Beijing, People's Republic of China}
\author{Hua Feng}
\affiliation{State Key Laboratory of Particle Astrophysics, Institute of High Energy Physics, Chinese Academy of Sciences, 100049 Beijing, People's Republic of China}
\author{Ming-Yu Ge}
\affiliation{State Key Laboratory of Particle Astrophysics, Institute of High Energy Physics, Chinese Academy of Sciences, 100049 Beijing, People's Republic of China}
\author{Lian Tao}
\affiliation{State Key Laboratory of Particle Astrophysics, Institute of High Energy Physics, Chinese Academy of Sciences, 100049 Beijing, People's Republic of China}
\author{Jing-Qiang Peng}
\affiliation{State Key Laboratory of Particle Astrophysics, Institute of High Energy Physics, Chinese Academy of Sciences, 100049 Beijing, People's Republic of China}
\affiliation{University of Chinese Academy of Sciences, Chinese Academy of Sciences, 100049 Beijing, People's Republic of China}
\author{Bo-Yan Chen}
\affiliation{State Key Laboratory of Particle Astrophysics, Institute of High Energy Physics, Chinese Academy of Sciences, 100049 Beijing, People's Republic of China}
\affiliation{University of Chinese Academy of Sciences, Chinese Academy of Sciences, 100049 Beijing, People's Republic of China}
\author{Ling-Da Kong}
\affiliation{Institut für Astronomie und Astrophysik, Kepler Center for Astro and Particle Physics, Eberhard Karls, Universität, Sand 1, D-72076 Tübingen, Germany}
\author{Peng-Ju Wang}
\affiliation{Institut für Astronomie und Astrophysik, Kepler Center for Astro and Particle Physics, Eberhard Karls, Universität, Sand 1, D-72076 Tübingen, Germany}

\begin{abstract}
We conduct an in-depth study of the quasi-periodic oscillation (QPO) properties of RE J1034+396, by constructing QPO phase-folded light curves from 10 XMM-Newton observations during 2020--2021. Our analysis reveals that the QPO in the source exhibits two mutually convertible lag-energy modes: ``hard lag" and ``soft lag". Despite different lag characteristics, the energy dependency of the root mean square (RMS) amplitude of the QPO under both modes are consistent, suggesting the two types of QPO originate from the same physical mechanism. By performing a spectral analysis, we further find a correlation between time-lag modes and spectral states: the soft lag mode typically corresponds to harder X-ray spectra and higher blackbody temperatures. Through comprehensive comparison of multiple theoretical models, we propose that the relativistic precession model (RPM) of the corona \textbf{\textbf{provides a plausible qualitative explanation for}} the observed complex phenomena, including time-lag mode transitions, and variations of spectral hardness and QPO signal strength.

\end{abstract}



\section{Introduction} \label{sec1}
Quasi-periodic oscillations (QPOs) are among the most important time-variable phenomena in high-energy astrophysics, widely observed in black hole X-ray binary systems (BHBs) and considered as important diagnostic tools for probing accretion flow dynamics in strong gravitational field environments. The discovery of QPO phenomena can be traced back to the \(1970\)s, when they were first observed in neutron star and black hole binary systems such as GX 339--4 \citep{1979Natur.278..434S}. However, although QPOs have been extensively studied in stellar-mass black hole systems, confirmed discoveries in supermassive black hole systems---active galactic nuclei (AGN)---remain extremely rare. \textbf{This disparity is primarily attributed to the intrinsic difference in QPO timescales between these two classes of sources, which naturally introduces a significant observational selection effect. Specifically, QPO periods in X-ray binaries typically range from milliseconds to hundreds of seconds, whereas those in AGNs are orders of magnitude longer. Consequently, signals with shorter periods are much easier to reliably detect within typical observational exposures, whereas the long-period QPOs characteristic of AGNs necessitate significantly longer durations of continuous monitoring. }

RE~J1034+396 is a narrow-line Seyfert~1 galaxy at redshift $z \approx 0.042$\citep{1995MNRAS.276...20P}. According to previous studies, its mass accretion rate approaches the Eddington limit \citep{2016A&A...594A.102C}, with the black hole mass between \(10^6\,M_{\odot}\) and \(10^7\,M_{\odot}\) \citep{2008Natur.455..369G,2009MNRAS.394..250M,2010MNRAS.401..507B,2012MNRAS.420.1825J,2016A&A...594A.102C,2018MNRAS.478.4830C,2021MNRAS.500.2475J}. \citet{2008Natur.455..369G} first discovered significant QPO signals in this source, with a characteristic frequency of approximately \(2.7 \times 10^{-4}\,\mathrm{Hz}\). This marked the first confirmed detection of QPO phenomena in AGNs, representing a milestone in QPO research. This discovery not only confirmed that supermassive black holes can also produce time-variable behavior similar to stellar-mass black holes, but also opened a new observational window for studying the physical mechanisms of AGN central engines.

QPO time lag studies are powerful tools for understanding the internal structure and radiation mechanisms of accretion flows \citep[see e.g.][]{2022NatAs...6..577M}. By analyzing time lag characteristics between radiation in different energy bands, one can reveal information about the spatial distribution of radiating regions, energy transport processes, and geometric structures \citep{2014A&ARv..22...72U}. In black hole X-ray binaries, QPO time lag studies have achieved rich results, providing important clues for understanding the interaction between accretion disk-corona systems \citep{2000ApJ...541..883R,2013ApJ...778..136P,2020MNRAS.497.4222C}. 

In the \(2007\) observation (observation ID 0506440101 of \textit{XMM-Newton}), for the first time, QPO signals were detected in RE J1034+396, which exhibited soft lag phenomena \citep{2008Natur.455..369G,2009MNRAS.394..250M}. \citet{2014ApJ...788...31H} pointed out that due to the varying QPO period, the previously discovered soft lag might be an artificial effect produced by folding light curves with a fixed period rather than a real physical process. They demonstrated this by folding the dataset using phases obtained from the Hilbert-Huang Transform (HHT), which made the phase lag disappear. In the \(2018\) observation (observation ID 0824030101), \citet{2020MNRAS.495.3538J} first discovered phase lag reversal (hard lag) between soft X-ray (\(0.3\)--\(1\,\mathrm{keV}\)) and hard X-ray (\(1\)--\(4\,\mathrm{keV}\)) light curves near the QPO frequency with high coherence, through ensemble empirical mode decomposition (EEMD) analysis. Subsequently, in a batch of observations from 2020--2021, \citet{2024ApJ...961L..32X} simultaneously observed both lag modes with high coherence, confirming that the QPO in RE~J1034+396 should exhibit both hard lag and soft lag (with mutual evolution between the two lag modes).

Regarding QPO phenomena in RE~J1034+396, various models have been established for discussion. \citet{2010ApJ...718..551M} proposed that QPOs are produced by periodic  obscuration from warm clouds, while \citet{2025ApJ...987..135T} considered that QPOs originate entirely from the corona. Additionally, \citet{2025ApJ...989...59X} proposed that QPOs originate from the accretion disk and are modulated by relativistic precession of the corona. Different time lags often correspond to different physical mechanisms \citep{2014ApJ...788...31H}, so careful study of the time lag characteristics of QPOs in RE~J1034+396 is crucial for distinguishing between these competing theoretical models. In recent years, with advances in X-ray observation technology, particularly the development of high time-resolution detectors, a technical foundation has been provided for in-depth study of QPO time lag characteristics in AGN \citep{2015MNRAS.449..467A,2015eheu.conf...76W,2019MNRAS.489.1957W,2023A&A...674A..83P,2024A&A...691A.252L,2024A&A...691A...7Y}. However, due to the relatively weak and intermittent nature of QPO signals in RE~J1034+396, existing time lag measurements still have large statistical errors and systematic uncertainties. Moreover, variations in QPO characteristics during different observation periods and multi-wavelength correlation studies also urgently need deeper exploration.

In this work, we aim to systematically study the time lag characteristics of QPOs in RE~J1034+396 by analyzing its long-term X-ray observational data \textbf{(\textbf{note:} all timescales involved in this article have been converted to the AGN rest frame, i.e., the timescales in the observed frame are divided by the time dilation factor $(1+z)$, where $z \approx 0.042$.}). In Section~\ref{sec2}, we mainly introduce the XMM-Newton satellite observational data and data processing methods; in Section~\ref{sec3}, we perform timeing analysis using HHT; in Section~\ref{sec4}, we conduct phase-averaged spectral analysis and discuss it in combination with the evolution of energy spectrum parameters; in Section~\ref{sec5}, we compare multiple theoretical models using our obtained results and provide the most likely model hypothesis currently available. Finally, we summarize our work in Section~\ref{sec6}.

\section{Observations and Data Reduction}\label{sec2}
The XMM-Newton satellite has conducted more than \(20\) observations of RE~J1034+396 since \(2002\) \citep{2008Natur.455..369G,2020MNRAS.495.3538J,2024ApJ...961L..32X}, accumulating a relatively rich observational sample. In this study, we selected the data from \(10\) publicly available observations by XMM-Newton \citep{2001A&A...365L...1J} during the period from November \(2020\) to June \(2021\) \citep{2024ApJ...961L..32X}. Each observation has a duration of approximately \(90\,\mathrm{ks}\).

\begin{deluxetable}{lcccc}
\tabletypesize{\scriptsize}
\tablewidth{0pt}
\tablecaption{XMM-Newton observations of RE J1034+396 and corresponding hardness ratios}
\tablehead{
  \colhead{Obs. No.} & \colhead{ObsID} & \colhead{Obs. Date } & \colhead{GTI(ks)} & \colhead{HR(*10\textsuperscript{\textsuperscript{-2}})}
}
\startdata
Obs1             & 0865010101  & 2020-11-20   & 86   & 13.33$^{+0.09}_{-0.09}$   \\
Obs2   & 0865011001                        & 2020-11-30
    & 85                         & 12.78$^{+0.08}_{-0.08}$   \\
Obs3        & 0865011101             & 2020-12-04
   & 91   & $12.80$$^{+0.08}_{-0.08}$ \\
Obs4      & 0865011201   & 2020-12-02   & 89     & 13.13$^{+0.08}_{-0.08}$     \\
Obs5    & 0865011301         & 2021-04-24         & 91      & 12.61$^{+0.08}_{-0.08}$     \\
Obs6     & 0865011401                         & 2021-05-02                         & 87    & 12.42$^{+0.08}_{-0.08}$   \\
Obs7   & 0865011501                        & 2021-05-08                         & 91          & 12.49$^{+0.08}_{-0.08}$         \\
Obs8       & 0865011601  & 2021-05-12  & 89  & 12.32$^{+0.08}_{-0.08}$  \\
Obs9     & 0865011701      & 2021-05-16      & 92     & 12.31$^{+0.08}_{-0.08}$     \\
Obs10     & 0865011801      & 2021-05-30      & 86     & 12.81$^{+0.08}_{-0.08}$     \\
\enddata
\tablecomments{Column definitions: (1) observation number assigned in this study; (2) observation ID; (3) observation date; (4) net exposure time; (5) hardness ratio and its error.}

\label{table1}
\end{deluxetable}

We used XMM-Newton Science Analysis Software SAS (version \(19.1.0\)) and the latest calibration files for data processing. Calibrated EPIC-PN event files were generated from the original observation data files through the \texttt{epproc} task. We extracted event files following standard data processing procedures and selected events with \(\mathrm{PATTERN} \leq 4\). For light curve extraction, following the selection method of \citet{2024ApJ...961L..32X}, we selected the source region within a circle centered on the source of interest with a radius of \(40\,\mathrm{arcseconds}\), while the background region was selected as a nearby source-free region of the same size. Considering the existence of gaps in the filtered data, we used the original event data (i.e., without GTI filtering on the original data) when generating light curves to better perform timing analysis. The extraction of source and background light curves was performed through the \texttt{evselect} task, and the background-subtracted light curves after the above region selection were generated by the \texttt{epiclccorr} task.

For timing analysis, we extracted light curves with time resolution of \(100\,\mathrm{s}\) (much smaller than the quasi-period of approximately \(3730\,\mathrm{s}\) \citep{2025ApJ...989...59X} )  in the \(0.3\)--\(10\,\mathrm{keV}\) energy range. Notably, for Obs1, Obs2, Obs9 and Obs10, we found that their QPO signals in the \(0.3\)--\(10\,\mathrm{keV}\) energy band are not significant, while the QPO signals  in the \(1\)--\(10\,\mathrm{keV}\) energy band remain significant, which is consistent with the results of \citet{2024ApJ...961L..32X}. Therefore, we presented light curves from these four observations in the 1--10\(\,\mathrm{keV}\) energy band to apparantly display the QPO features.

For the spectral analysis, to avoid the influence of pile-up effects \citep{2009MNRAS.394..250M,2010ApJ...718..551M}, we uniformly excluded a circular region of \(10\,\mathrm{arcseconds}\) centered on the source position for all \(10\) observations, and extracted events from an annular region with inner and outer radius of \(10\) and \(20\,\mathrm{arcseconds}\) respectively. The background was extracted from a nearby \(40\,\mathrm{arcsecond}\) radius circular region free of other sources. We used the \texttt{evselect} task to extract spectra from the selected regions, applied the \texttt{backscale} task for effective area calibration, used \texttt{rmfgen} and \texttt{arfgen} tasks to create response matrices and auxiliary response files, and finally used the \texttt{specgroup} tool to group the spectra, to ensure that each spectral channel has at least \(25\) counts while the sampling factor of the instrumental energy resolution does not exceed \(3\). This enables the application of \(\chi^2\) statistics in spectral modeling.

\section{Timing Analysis} \label{sec3}
\subsection{QPO Search}\label{sec3.1}

\begin{figure*}[!htbp]
    \centering
    \begin{minipage}{0.46\textwidth}
        \centering
        \includegraphics[width=\textwidth]{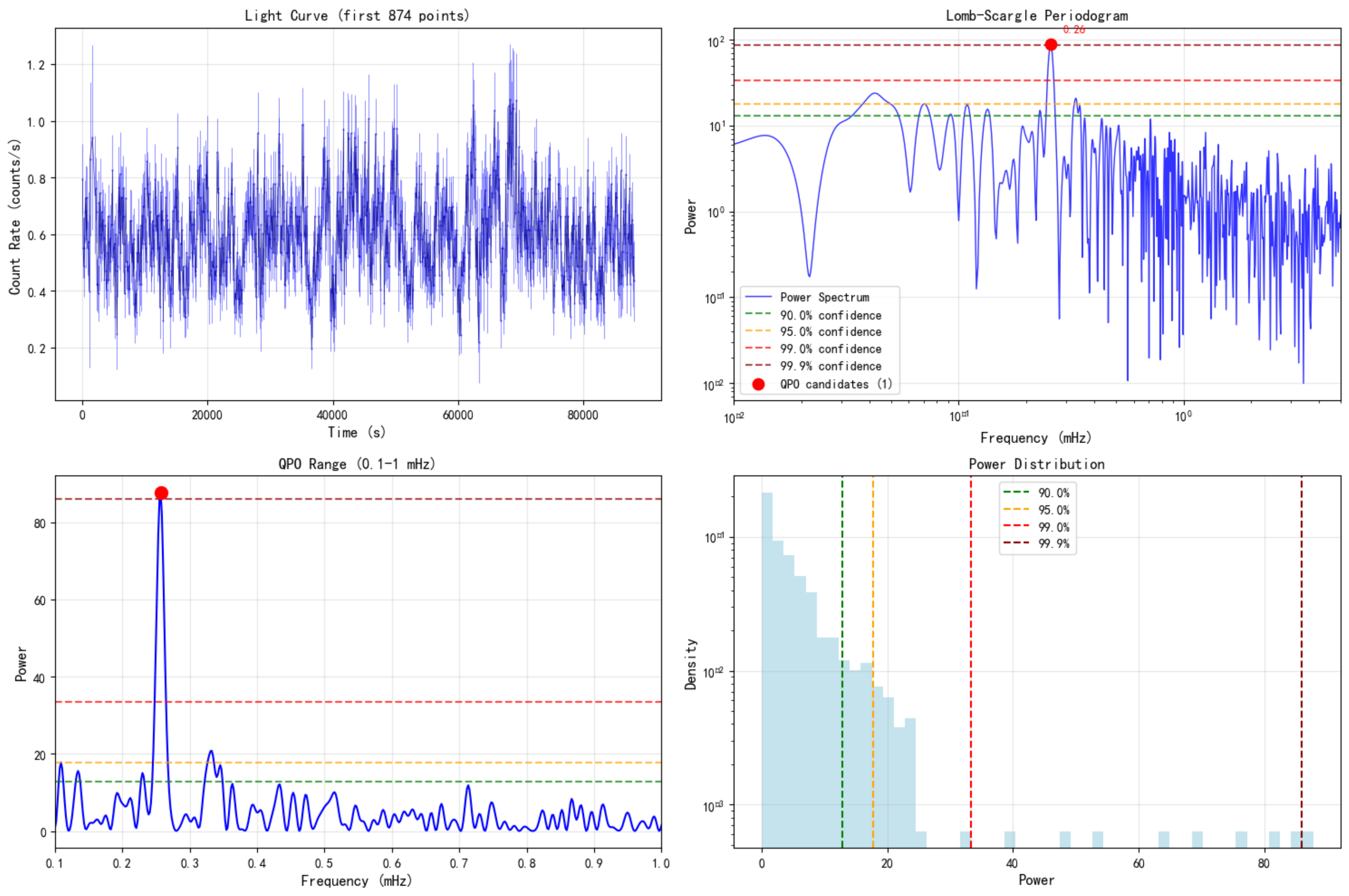}
        \vspace{0.2cm}  
        \makebox[0pt][c]{\textbf{(a)}}
    \end{minipage}
    \begin{minipage}{0.46\textwidth}
        \centering
        \includegraphics[width=\textwidth]{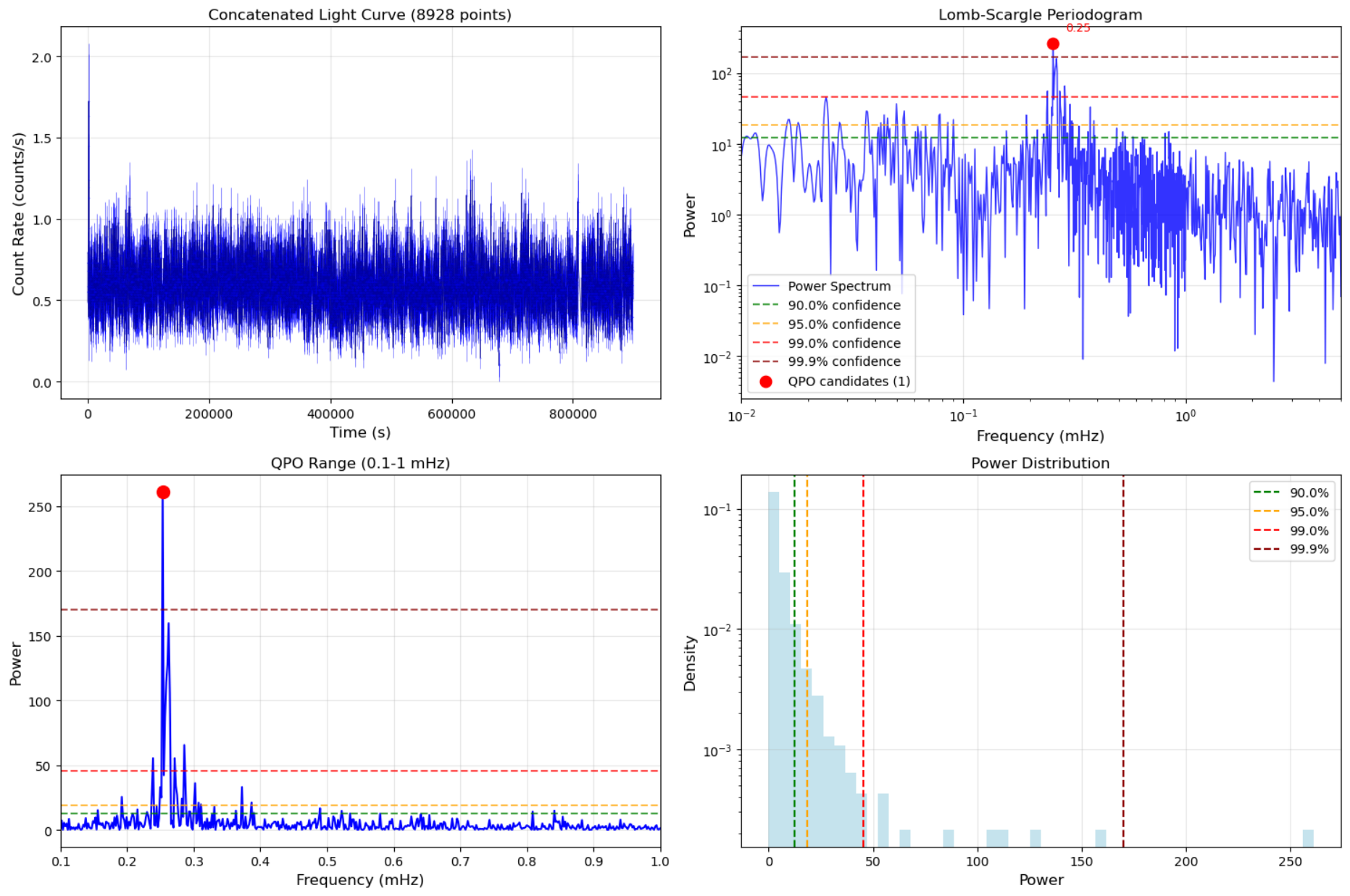}
        \vspace{0.2cm}  
        \makebox[0pt][c]{\textbf{(b)}}
    \end{minipage}
    \vspace{0.2cm}  

    \caption{%
         (a),(b)Lomb-Scargle periodogram analysis of observation 1 and combined data from 10 observations (1--10 keV). Upper left: light curve; Upper right: power spectrum over 0.01-5 mHz; Lower left: detailed power spectrum in the QPO frequency band (0.1-1 mHz); Lower right: power distribution. The dashed line shows the statistical significance level. 
         }
    \label{fig1}
\end{figure*}

\textbf{To accurately extract QPO features from the light curves, we first employed conventional time series analysis methods--Lomb-Scargle periodogram (LSP) to establish the presence and basic characteristics of QPO signals in the XMM-Newton observational data. For specific results, taking observation 1 as an example, Figure~\ref{fig1}a respectively shows its light curve, broad-frequency power spectrum, power spectrum near QPO, and power distribution. It can be seen that the red noise component contributes to the power at low frequencies, but the QPO signal remains significant. Additionally, we found that the QPO frequencies of these 10 observations are relatively close, so we combined the light curves (1--10 keV) from the 10 observations and analyzed their average period (results shown in Figure~\ref{fig1}b). The QPO signal was further enhanced (approximately 3950s, converted to approximately 3790s in the AGN rest frame).
}

\textbf{Furthermore, we adopted the Hilbert-Huang Transform analysis procedure described by \citet{2025ApJ...995L..30S} to perform phase-resolved analysis of QPOs. }

\textbf{HHT is an adaptive data analysis technique originally proposed by \citet{1998RSPSA.454..903H} for analyzing nonlinear and non-stationary signals. Unlike traditional Fourier analysis, which assumes signal linearity and stationarity, HHT can effectively analyze complex signals whose frequency and amplitude vary with time, making it particularly suitable for studying QPOs in astrophysical sources—oscillations whose frequency and phase evolve over time.}

\textbf{The HHT method consists of two core components: (1) mode decomposition, which decomposes complex signals into several Intrinsic Mode Functions (IMFs), and (2) Hilbert Spectral Analysis (HSA), which extracts instantaneous frequency, phase, and amplitude information for each IMF. In QPO studies, the key advantage of this method lies in its ability to track the time-varying characteristics of QPO frequencies and provide precise instantaneous phase information for phase-resolved analysis. Traditional period-folding methods struggle to achieve this goal due to the inherent non-deterministic nature of QPO phase evolution \citep{1997ApJ...482..993M}.}

\textbf{In recent years, applications of HHT in astrophysics have demonstrated effectiveness across multiple fields, including pulsar timing studies \citep{2011ApJ...740...67H}, solar activity analysis \    \citep{2011SoPh..269..439B}, and black hole binary QPO analysis \citep{2015ApJ...815...74S,2023ApJ...951..130Y,2023ApJ...957...84S,2025Natur.638..370M}. Particularly relevant to this work, \citet{2014ApJ...788...31H} first applied HHT to QPO analysis of RE J1034+396, demonstrating that this method can reliably extract QPO features from active galactic nucleus (AGN) light curves.}

\textbf{Based on recent advances, this study employs the state-of-the-art Variational Mode Extraction (VME) algorithm \citep{Nazari2018Variational}, which represents a significant improvement over traditional HHT methods. The core idea is to transform signal decomposition into a constrained optimization problem that can be directly solved in the frequency domain, achieving signal decomposition through direct mathematical optimization while avoiding artificially generated mode mixing artifacts. This approach can precisely target specific frequency components (in this study, the $\sim$2.7$\times$10$^{-4}$ Hz quasi-periodic oscillation signal) while reducing computational requirements (for specific details about vme, see Appendix B in \citet{2025ApJ...995L..30S}.}

\textbf{For each background-subtracted PN light curve, we applied VME to extract QPO modes, with algorithm parameters optimized for mHz-scale QPO extraction (central frequency $\approx$ 2.7$\times$10$^{-4}$ Hz). We used Hilbert Spectral Analysis (HSA) to obtain the instantaneous phase information of the signal \citep[referring to][]{2014ApJ...788...31H,2023ApJ...957...84S,2024ApJ...973...59S,2024ApJ...973...92S,2024ApJ...961L..42Z}. As an example, Figure~\ref{fig2}a presents the original source light curve (in blue), corresponding extracted intrinsic QPO light curve (in red) and instantaneous QPO phase (in orange) from Observation 5, clearly revealing the temporal evolution characteristics of the oscillation.}

\subsection{Energy-dependent Analysis}\label{sec3.2}
\begin{figure}[!htbp]
    \centering
    \begin{minipage}{0.47\textwidth}
        \centering
        \includegraphics[width=\textwidth]{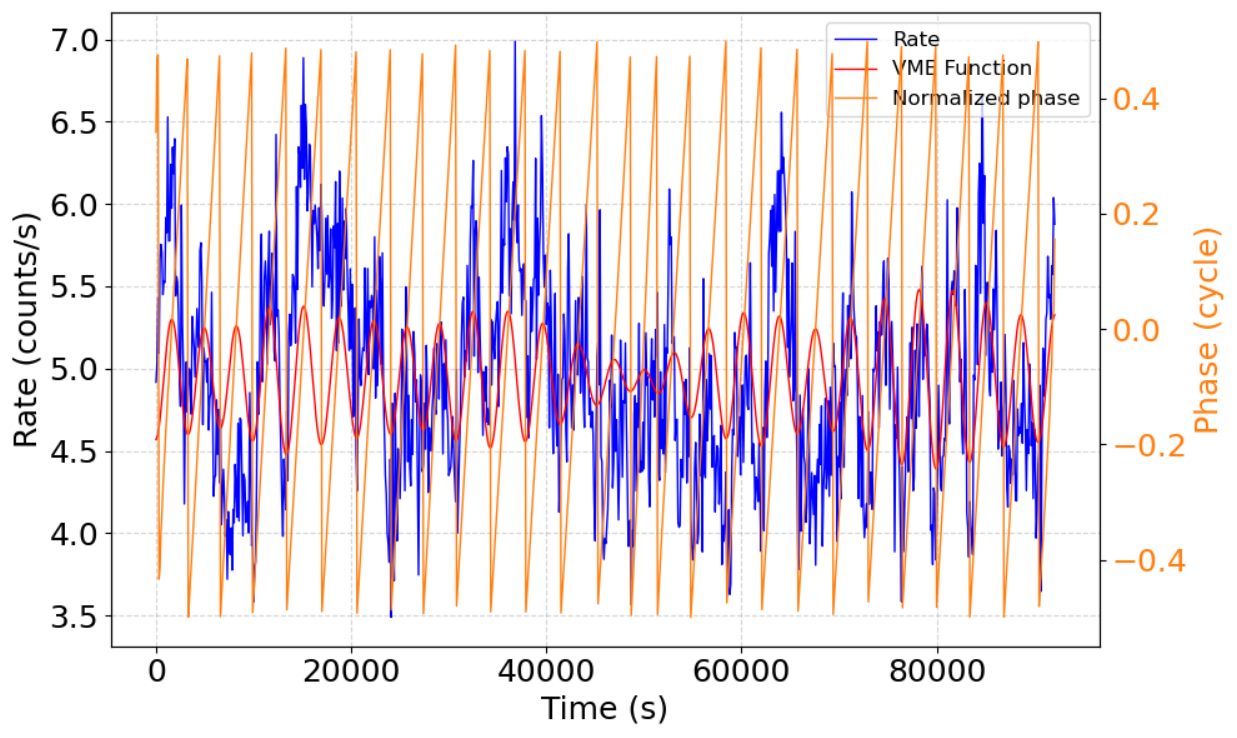}
        \vspace{0.2cm}  
        \makebox[0pt][c]{\textbf{(a)}}
    \end{minipage}
    \hspace{0.1in}
    \begin{minipage}{0.45\textwidth}
        \centering
        \includegraphics[width=\textwidth]{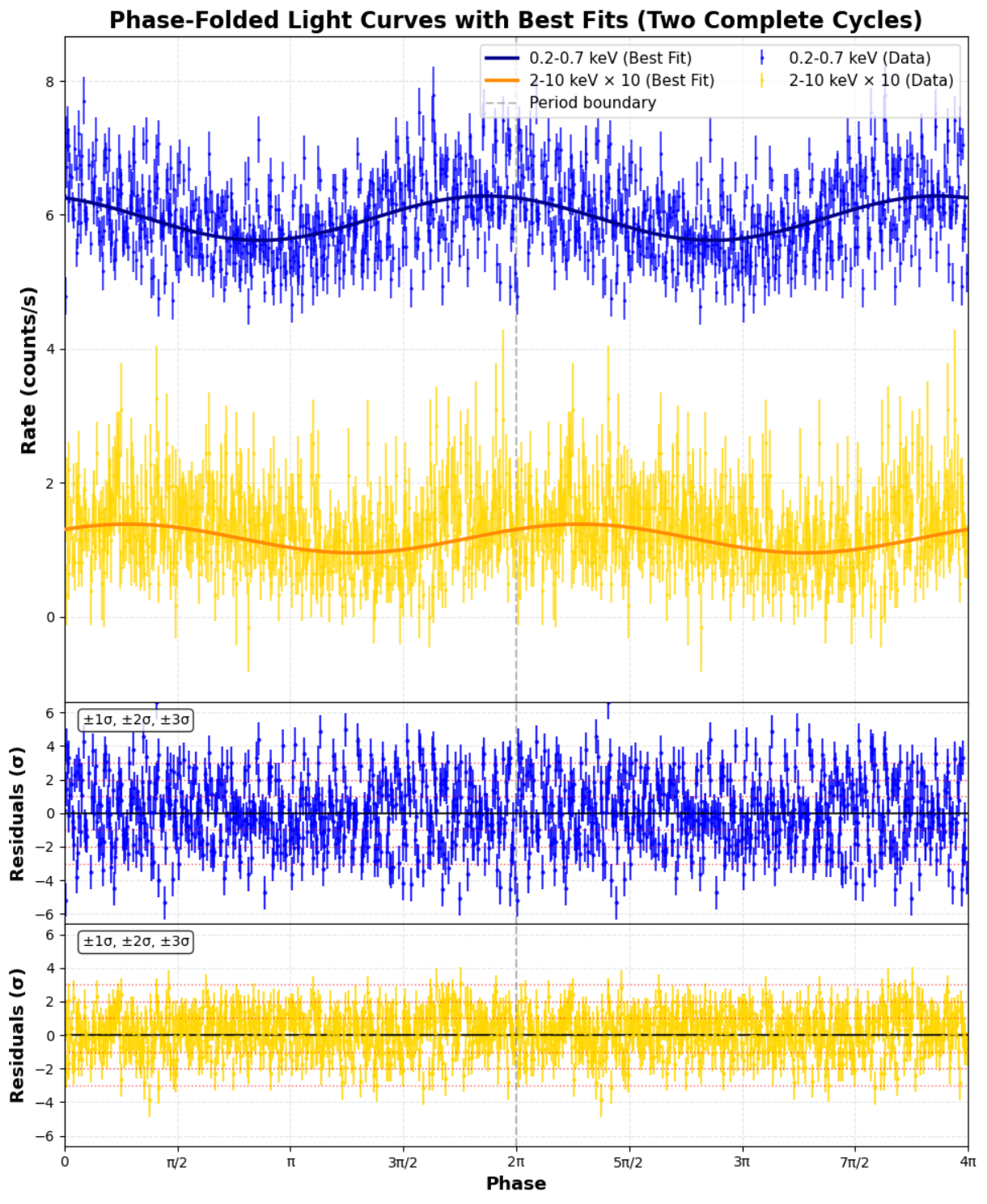}
        \vspace{0.2cm}  
        \makebox[0pt][c]{\textbf{(b)}}
    \end{minipage}
    \hspace{0.1in}
    \begin{minipage}{0.45\textwidth}
        \centering
        \includegraphics[width=\textwidth]{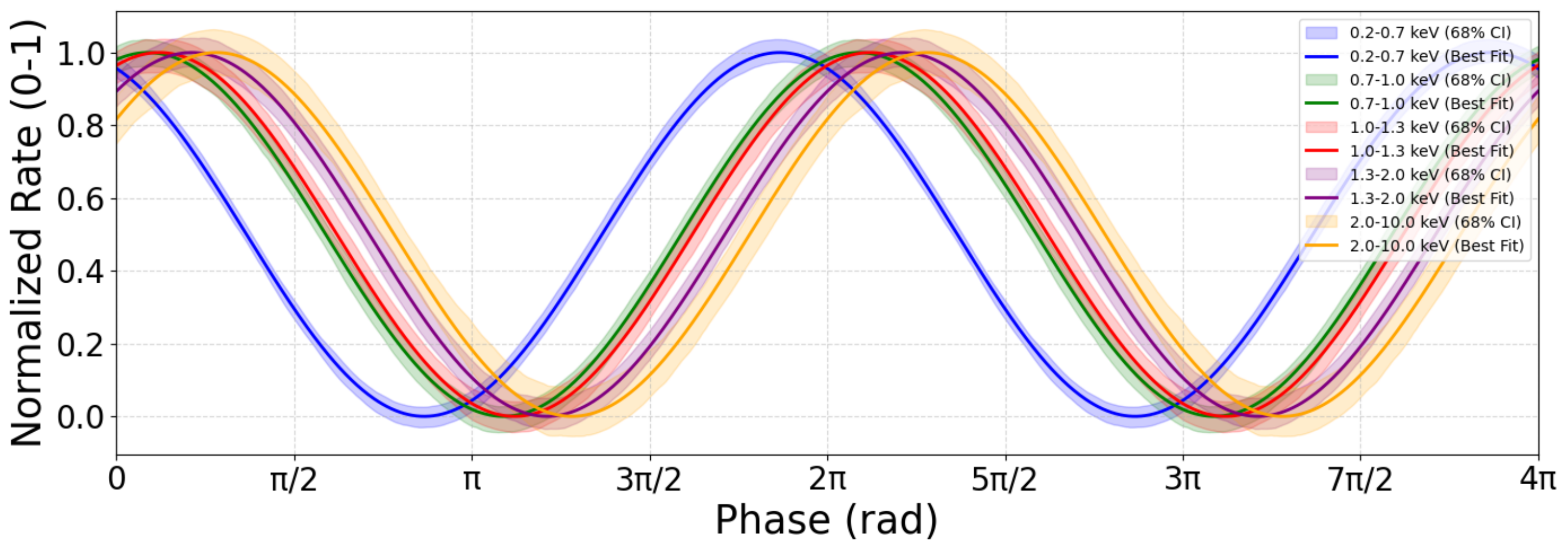}
        \vspace{0.2cm}  
        \makebox[0pt][c]{\textbf{(c)}}
    \end{minipage}
    \hspace{0.1in}
    \vspace{0.2cm}  

    \caption{%
         (a) Light curve of Observation 5 analyzed using Hilbert-Huang Transform. Blue line represents the original light curve (with 100-second time binning), red line represents the intrinsic QPO light curve, and orange line represents the instantaneous QPO phase obtained using HSA; (b) MCMC fitting results and residuals for two-period phase-folded light curves of observation 5 in the 0.2--0.7 keV (blue) and 2--10 keV (yellow, count rate multiplied by 10) energy bands. Points and short lines represent binned data points with errors, and long lines represent MCMC fitted curves (with width representing 1$\sigma$ confidence intervals). (c) Normalized fitting curves and their confidence intervals for Observation 5 across five energy bands: 0.2--0.7\(\,\mathrm{keV}\), 0.7--1\(\,\mathrm{keV}\), 1--1.3\(\,\mathrm{keV}\), 1.3--2\(\,\mathrm{keV}\), and 2--10\(\,\mathrm{keV}\).
  }
    \label{fig2}
\end{figure}

\textbf{To investigate the energy dependence of QPOs, we performed energy binning of the light curves. Considering the count rates and XMM-Newton's larger effective area around 1\(\,\mathrm{keV}\) (poor signal-to-noise ratios at low count rates can significantly affect the accuracy and reliability of fitting results), we employed finer energy binning in the lower energy bands and around 1\(\,\mathrm{keV}\), while selecting broader energy intervals in the higher energy range. After testing various configurations, we ultimately selected five energy bands: 0.2--0.7\(\,\mathrm{keV}\), 0.7--1\(\,\mathrm{keV}\), 1--1.3\(\,\mathrm{keV}\), 1.3--2\(\,\mathrm{keV}\), and 2--10\(\,\mathrm{keV}\) for phase folding of the light curves.} For the phase-folded light curves, we performed fitting using the Monte Carlo Markov Chain (MCMC) method. We used the mean count rate $\mu$, amplitude $A$ of the phase-folded light curve, and phase shift $\phi$ as parameters, with the model function being a sine function: $\mu + A \times \sin(x + \phi)$, and constructed a log-likelihood function incorporating data errors \citep[referring to][]{2020MNRAS.495.3538J}. We used the median of the parameter posterior distributions as fitted values and calculated parameter confidence intervals using the 16$^{\mathrm{th}}$ and 84$^{\mathrm{th}}$ percentiles \citep[referring to][]{2024ApJ...973...59S}, while also calculating the percentage of $A/\sqrt{2}\mu$ and its error. Finally, we obtained parameters including amplitude ($A$), phase shift ($\phi$), mean count rate ($\mu$), and fractional RMS (the root-mean-square fractional amplitude of QPO, calculated as $A/\sqrt{2}\mu$). \textbf{Figure~\ref{fig2}b shows the phase-folded light curves and corresponding fitting results for observation 5 in two energy bands: 0.2--0.7 keV and 2--10 keV. To more intuitively display the phase variations across different energy bands for the same observation, we plot together the normalized fitting curves (with the minimum point of the fitting curve as 0 and the maximum point as 1) and confidence intervals for observation 5 across 5 energy intervals, as shown in Figure~\ref{fig2}c.}

\begin{figure*}[!htbp]
    \centering
    \begin{minipage}{1\textwidth}
        \centering
        \includegraphics[width=\textwidth]{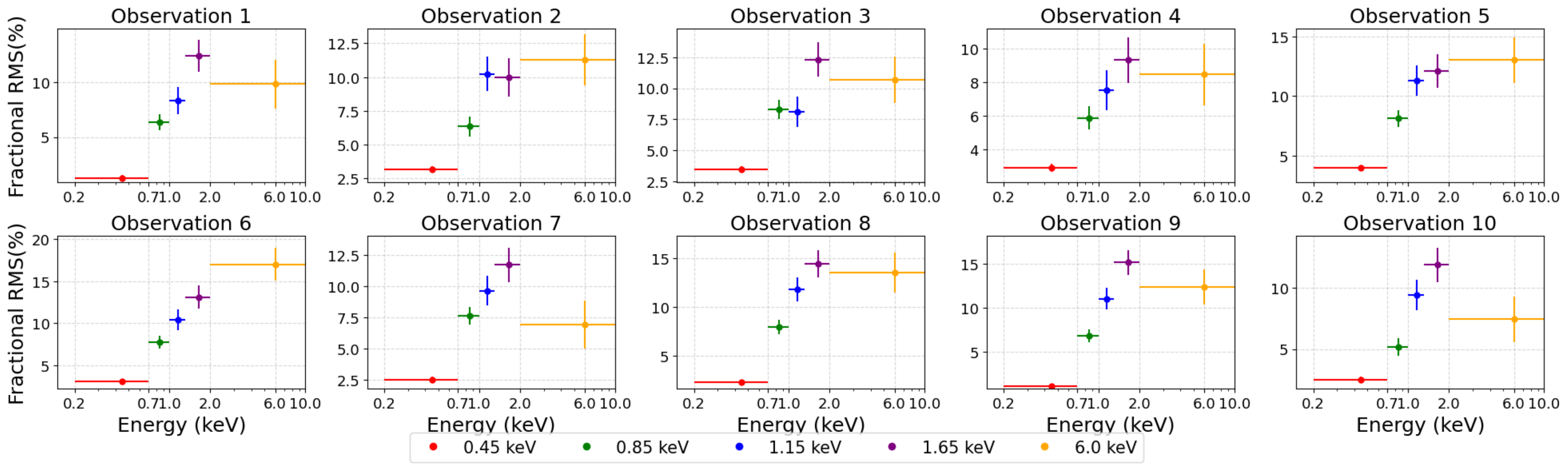}
        \vspace{0.2cm}  
        \makebox[0pt][c]{\textbf{(a)}}
    \end{minipage}
    \hspace{0.1in}
    \begin{minipage}{1\textwidth}
        \centering
        \includegraphics[width=\textwidth]{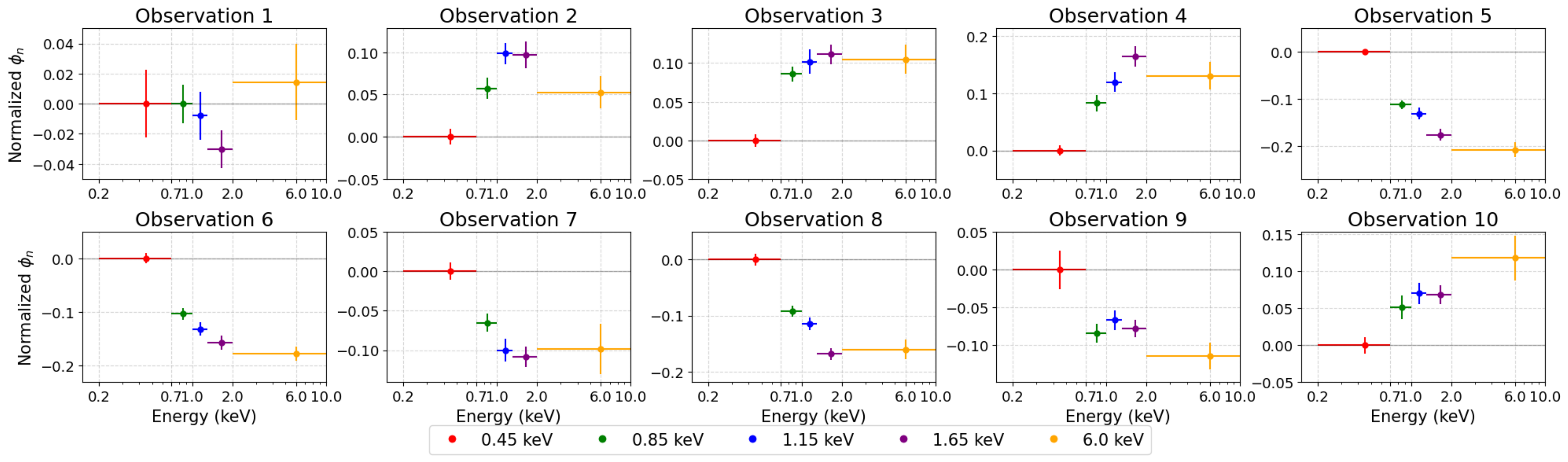}
        \vspace{0.2cm}  
        \makebox[0pt][c]{\textbf{(b)}}
    \end{minipage}
    \hspace{0.1in}
    \begin{minipage}{0.48\textwidth}
        \centering
        \includegraphics[width=\textwidth]{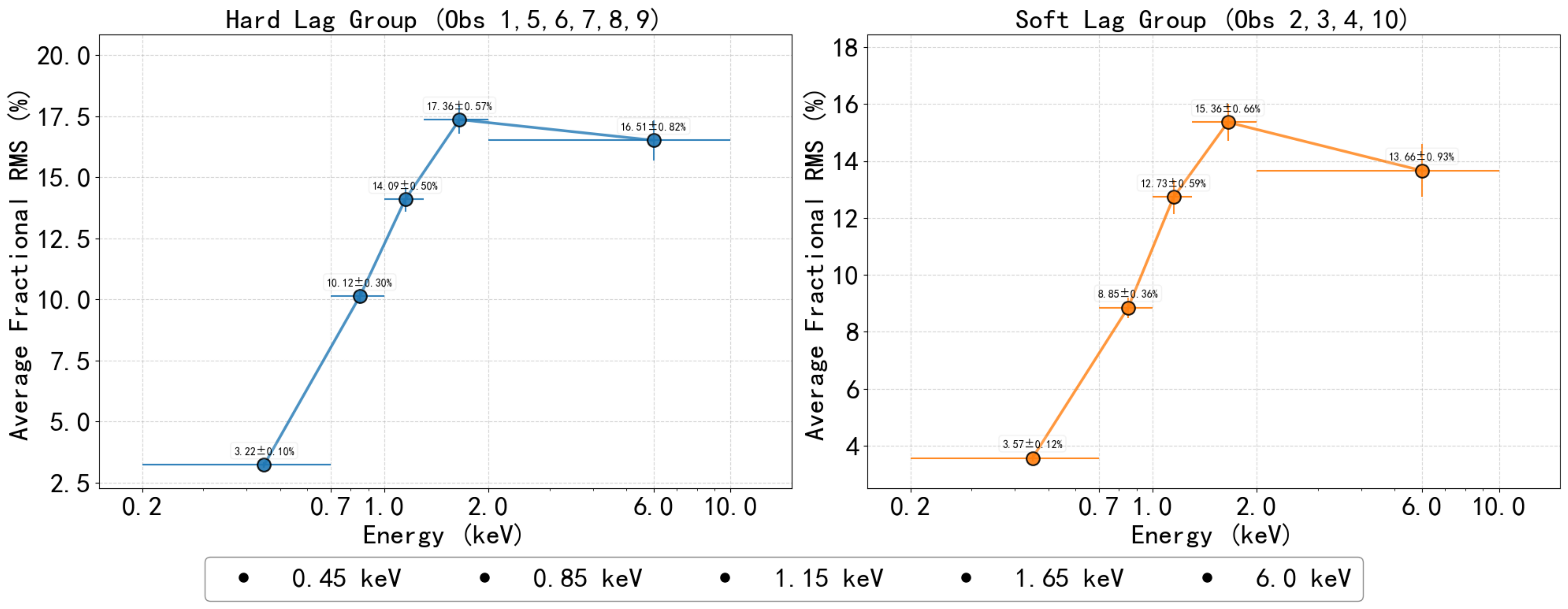}
        \vspace{0.2cm}  
        \makebox[0pt][c]{\textbf{(c)}}
    \end{minipage}
    \hspace{0.1in}
    \begin{minipage}{0.48\textwidth}
        \centering
        \includegraphics[width=\textwidth]{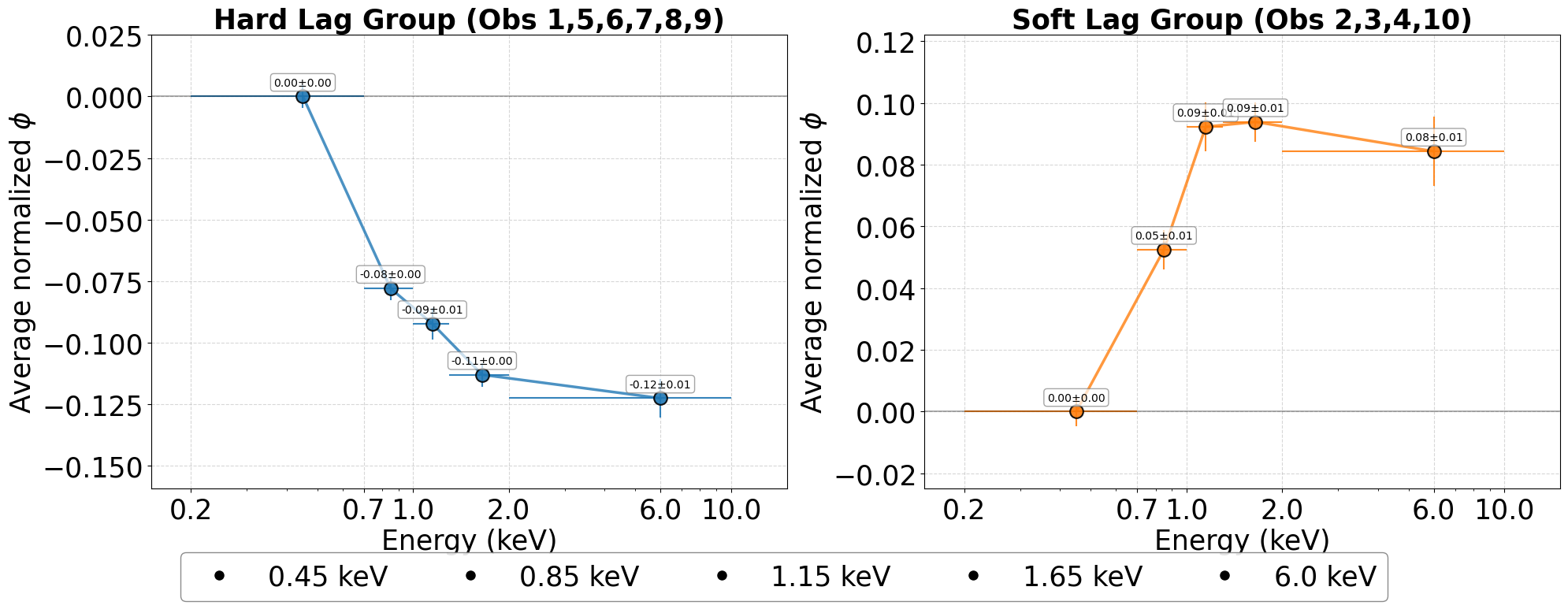}
        \vspace{0.2cm} 
        \makebox[0pt][c]{\textbf{(d)}}
    \end{minipage}
    \hspace{0.1in}

    \caption{%
       (a) Fractional rms of QPO versus energy for \(10\) observations; (b) Energy dependence of the phase lag for \(10\) observations; (c) , (d) Fractional rms-energy and phase shift-energy relationship after combining soft and hard lag data. Both fractional rms and phase lags are from the best-fitting results of MCMC.
}
    \label{fig3}
\end{figure*}

We plot the relationship between the fractional rms of QPOs and energy for the \(10\) observations in Figure~\ref{fig3}a. \textbf{Overall, the fractional RMS exhibits an increasing trend in the 0.2--2\(\,\mathrm{keV}\) range. At high-energy X-rays ($>$2\(\,\mathrm{keV}\)), different observations show varying behaviors (most observations display relatively flat RMS variations with energy, while some observations exhibit continued rise or decline with increasing energy, which may be influenced to some extent by statistical uncertainties arising from low count rates in the high-energy band).} It is a typical feature reported by previous studies \citep[e.g.][]{10.1111/j.1745-3933.2009.00697.x,2013MNRAS.436.3173J,2016MNRAS.455..691J,2017MNRAS.468.3663J}, indicating that the fractional variation in hard X-rays is larger than that in soft X-rays. Phase lags among different bands are an intriguing issue in AGNs, and different lag scales may correspond to different physical mechanisms \citep{2014ApJ...788...31H}. \textbf{Figure~\ref{fig3}b shows the energy dependence of phase lag, where the reference band was set to 0.2--0.7\(\,\mathrm{keV}\) (we normalized the fitted phase shift $\phi$ by $\phi/(2\pi)$ and used the result from the 0.2--0.7\(\,\mathrm{keV}\) band as the reference zero point, i.e., normalized $\phi= \phi/(2\pi) - \phi_{0.2-0.7}/(2\pi)$)}. It can be seen that the phase shifts of observations \(5\), \(6\), \(7\), \(8\), \(9\) decrease with energy, while those of observations \(2\), \(3\), \(4\), \(10\) increase with energy. The phase shift of observation 1 shows no obvious variation with energy, but a decreasing trend with increasing energy can still be seen in several data points at lower energy bands (\(0.2\)--\(2\,\mathrm{keV}\)), so we classify it with observations \(5\)--\(9\). \textbf{In the MCMC fitting function $\mu + A \times \sin(x + \phi)$, $\mu$, $A$, and $\phi$ are parameter values obtained from the fitting, while $x$ is directly associated with the photon arrival time $t$, representing the corresponding phase value of photon arrival time $t$ after the Hilbert-Huang Transform, which is positively correlated with $t$ within one period. When $\mu + A \times \sin(x + \phi)$ corresponds to a specific position in the QPO sinusoidal waveform, $x$ and $\phi$ can be considered anti-correlated (for example, when corresponding to the peak, $x + \phi$ equals $\pi/2$ (or $\pi/2$ plus some integer multiple of $2\pi$)). In this case, a larger $\phi$ corresponds to a smaller $x$, indicating a smaller photon arrival time, i.e., photons arrive earlier. Observations 2, 3, 4, and 10 show larger $\phi$ values in the hard energy bands, indicating that their hard photons arrive earlier, thus exhibiting soft lags. Conversely, Observations 5, 6, 7, 8, and 9 exhibit hard lags.} This is consistent with the results of \citet{2024ApJ...961L..32X}. To improve statistics and examine the differences between the two lag modes, we designated observations \(1\), \(5\), \(6\), \(7\), \(8\), \(9\) as the hard lag group and observations \(2\), \(3\), \(4\), \(10\) as the soft lag group. Then we combined the data from the two groups, re-plotted light curves following the above process, extracted phases and performed phase folding by energy bands. Fianlly, we obtained fractional rms and phase shifts through MCMC fitting. The fractional rms-energy relationship and phase lag-energy relationship plots for the combined data from the two groups are shown in Figures~\ref{fig2}c and \ref{fig2}d. It can be seen that the fractional rms variations of soft and hard X-rays in two lag modes are consistent, suggesting that QPOs in these two lag modes may have the same origin. Additionally, from the relationship between phase lag and time lag (\(\Delta\tau = T\Delta\phi /2\pi\), where \(T\) is the QPO period, here we take \(3730\,\mathrm{s}\) \citep{2025ApJ...989...59X}), we can roughly calculate that the soft lag timescale is approximately \(430\)--\(575\,\mathrm{s}\), and the hard lag timescale is approximately \(430\)--\(720\,\mathrm{s}\) \textbf{(timescales have been converted to the AGN rest frame)}.

\begin{figure}[!htbp]
    \centering
    \begin{minipage}{0.45\textwidth}
        \centering
        \includegraphics[width=\textwidth]{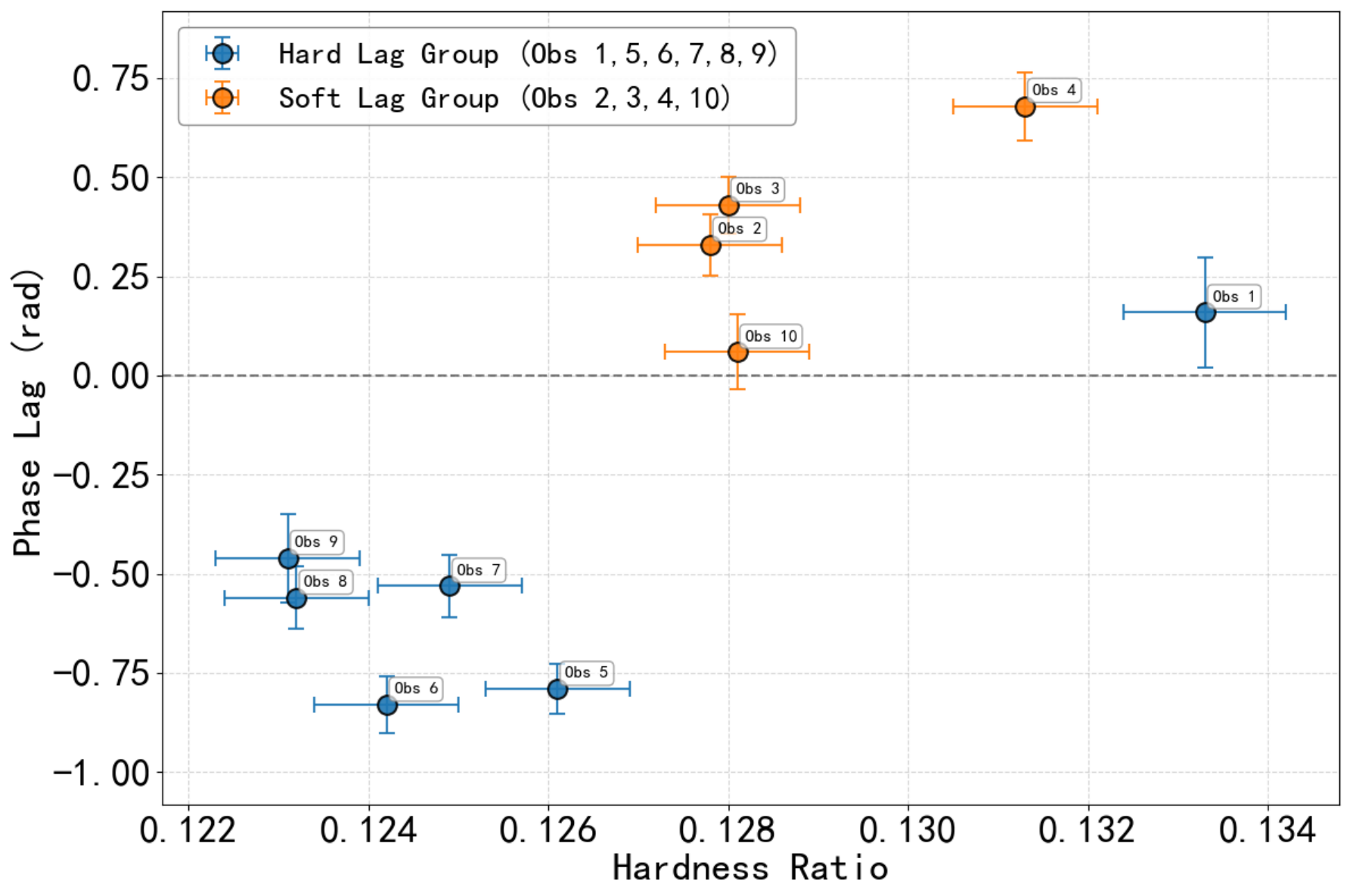}
        \vspace{0.2cm}  
    \end{minipage}
    \hspace{0.1in}

    \caption{%
      Relationship between hardness ratio and phase lag for 10 observations, where hardness ratio $HR = H/S$, with $H$ being the count rate in the hard band (\(1\)--\(10\,\mathrm{keV}\)) and $S$ being the count rate in the soft band (\(0.3\)--\(1\,\mathrm{keV}\)). Phase lag is taken as the difference between phase shift values of the \(1\)--\(10\,\mathrm{keV}\) and \(0.3\)--\(1\,\mathrm{keV}\) energy bands.}
    \label{fig4}
\end{figure}

Subsequently, we calculated the hardness ratio for the \(10\) observations using \(1\,\mathrm{keV}\) as the boundary between soft and hard energy bands (i.e., the ratio of average count rates in light curves between \(1\)--\(10\,\mathrm{keV}\) and \(0.3\)--\(1\,\mathrm{keV}\)), and performed phase folding of light curves with MCMC fitting for each observation in \(0.3\)--\(1\,\mathrm{keV}\) and \(1\)--\(10\,\mathrm{keV}\), obtaining phase shifts (\(\phi\)) in these two energy bands. All errors were calculated using error propagation formulas. The phase lag for each observation is taken as the difference between the phase shift values of the \(1\)--\(10\,\mathrm{keV}\) and \(0.3\)--\(1\,\mathrm{keV}\) energy bands. Positive values indicate larger phase shift values with smaller absolute photon arrival time in the high-energy band, corresponding to soft lag, while negative values correspond to hard lag. Figure~\ref{fig4} shows the relationship between hardness ratio and phase lag, and observations of the two lag mode groups are in different colors. It can be clearly seen that the hardness ratios of the soft lag group are significantly larger than those of the hard lag group, with observation 1 not being considered due to the peculiarity of its phase lag (see Figure~\ref{fig3}b). This indicates that the transition from hard lag to soft lag may be accompanied by spectral hardening. This trend is opposite to the results previously presented in black hole binary studies of type-B QPOs by \citet{2017MNRAS.466..564G}.

\section{Energy Spectral Analysis} \label{sec4}

\begin{figure*}[!htbp]
    \centering
    \begin{minipage}{0.7\textwidth}
        \centering
        \includegraphics[width=\textwidth]{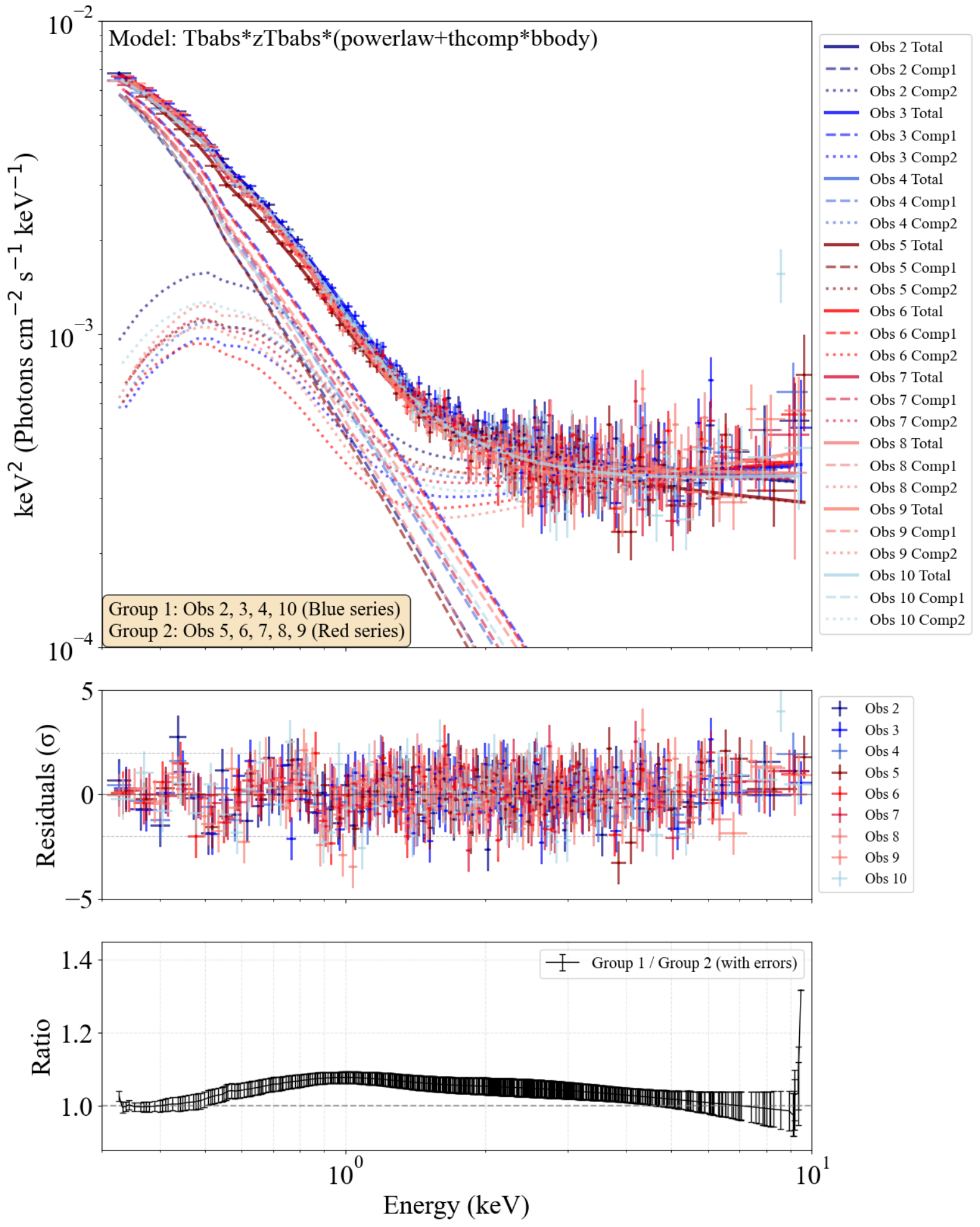}
        \vspace{0.1cm} 
    \end{minipage}
    \hspace{0.1in}

    \vspace{0.2cm}  

    \caption{%
       \textit{Upper panel:} Joint fitting results of energy spectra (eeuf) for \(9\) observations, with blue representing soft lag and red representing hard lag. Component 1 represents \texttt{powerlaw} and component 2 represents \texttt{thcomp*bbody};
\textit{Middle panel:} Residual plots of energy spectra for the \(9\) observations;
\textit{Bottom panel:} The variation of average flux ratio and its error between the soft lag group and hard lag group with energy, calculated from the total model.
}
    \label{fig5}
\end{figure*}

To further investigate the origin and differences between the two QPO phase-lag modes, we extracted EPIC-PN spectra from \(9\) observations (considering the peculiarity of observation 1 in timing analysis, we excluded it in spectral analysis), and performed spectral analysis using the \textsc{xspec} v\(12.14.0\) software package \citep{1996ASPC..101...17A}. We first attempted to fit the spectra using a power-law model (\texttt{powerlaw}) and a blackbody radiation model (\texttt{bbody}) as emission components, but found poor fitting results (\(\chi^2/\mathrm{d.o.f}\) is approximately between 4 and 5). Subsequently, we added a thermal Comptonization component and chose to use a powerlaw model along with a composite model combining thermal Comptonization and blackbody radiation as emission components, while using \texttt{tbabs} and \texttt{ztbabs} as absorption components to model the spectra,\textbf{ and employing the \texttt{cflux} model to obtain the flux of the \texttt{thcomp*bbody} component: \texttt{tbabs*ztbabs*(powerlaw+cflux*thcomp*bbody)}. }The \texttt{tbabs} model was used to model interstellar absorption \citep{2000ApJ...542..914W}, with the equivalent hydrogen column density parameter \(N_{\mathrm{H},\mathrm{gal}}\) fixed at \(1.31 \times 10^{20}\,\mathrm{cm^{-2}}\). The \texttt{ztbabs} model was used to model absorption around the AGN, with the redshift parameter fixed at \(0.042\). \textbf{Additionally, we used the \texttt{xset} command to set \texttt{POW\_EMIN} and \texttt{POW\_EMAX} to \(0.3\,\mathrm{keV}\) and 1\(\,\mathrm{keV}\) respectively, and set the minimum and maximum energies in \texttt{cflux} to 0.3\(\,\mathrm{keV}\) and 10\(\,\mathrm{keV}\) respectively. }In the thermal Comptonization model, we fixed the electron temperature parameter $kT_e$ at 150\(\,\mathrm{keV}\) and the redshift parameter at 0.042.\textbf{ In the blackbody model (bbody), we fixed the blackbody normalization factor ($N_{bb}$) to 1. Other parameters were allowed to vary freely. Considering the assumed 150\(\,\mathrm{keV}\) $kT_e$ value, we extended the energy range from 0.03\(\,\mathrm{keV}\) to 300\(\,\mathrm{keV}\) through the \texttt{energies} command and used 2000 energy bins on a logarithmic scale to achieve sufficient energy resolution for model ratio calculations (the bottom panel of Figure~\ref{fig5}) within the 0.3--10\(\,\mathrm{keV}\) band. Additionally, when plotting the spectra, we performed rebinning using the \texttt{setplot rebin} command (5$\sigma$ significance threshold, combining at most 10 bins each time).}

\textbf{After jointly fitting the average spectra of 9 observations using the above model, we ran MCMC chains to explore the parameter space and obtained the best-fit values and uncertainties for 7 free parameters at the 1$\sigma$ confidence level: self-absorption equivalent hydrogen column density ($N_{H,host}$), photon index ($\Gamma$), power-law flux ($F_{pl}$), Comptonization flux ($F_{th}$), thermal Comptonization photon power-law index (\(\Gamma_{\tau}\)), covering factor ($f_{cov}$), and blackbody temperature ($kT_{bb}$).} To further assess the convergence of the MCMC chains, we compared the one-dimensional and two-dimensional projections of the posterior distributions for each parameter between the first and second halves of the chains and found no significant differences. Taking Observation 2 as an example, we provide contour maps and probability distributions for each free parameter in the joint phase-resolved spectral analysis in the Appendix~\ref{appendix a}. The best-fit values and uncertainties are shown in Table~\ref{table2} along with the $\chi^2$/d.o.f. values, with a total $\chi^2$/d.o.f. of 1160.33/1051 for the joint fitting results of 9 observations. \textbf{The rebinned joint fitting results (\texttt{eeufspec}) are displayed in the upper panel of Figure~\ref{fig5}, with the red series representing the hard lag group (Observations 5, 6, 7, 8, 9) and the blue series representing the soft lag group (Observations 2, 3, 4, 10). } The middle panel shows the residual plot. Additionally, we did not find significant iron K\(\alpha\) emission lines in the spectra, consistent with previous results from \citet{2021MNRAS.500.2475J}. Subsequently, we averaged the total model flux of the soft lag group and calculated its ratio to the average total model flux of the hard lag group, and plotted the variation of this ratio with energy (the bottom panel of Figure~\ref{fig5}). \textbf{It can be seen that the average flux ratio between the soft lag group and hard lag group remains almost consistently above 1 (0.4--7\(\,\mathrm{keV}\)), and it first increases then decreases with increasing energy, reaching a peak around 1\(\,\mathrm{keV}\) (the significant rise in this ratio near 10\(\,\mathrm{keV}\) may be caused by factors such as signal-to-noise ratio and statistical uncertainties, as confirmed by the increasing error bars, indicating that the model ratio results become unreliable at higher energies). This demonstrates that, overall, the soft lag group exhibits slightly higher flux than the hard lag group, with a notable flux excess around 1\(\,\mathrm{keV}\) compared to the hard lag group. Furthermore, considering the hard excess in the 6--10\(\,\mathrm{keV}\) band observed in the spectra, we additionally included a reflection component (\texttt{pexriv}) in the fitting. But found that while the fitting performance in the high-energy band has improved(with the hard excess eliminated), the goodness of fit remained nearly unchanged (the $\chi^2$/d.o.f. variations were within 3\% for all observations, with most observations showing $\chi^2$/d.o.f. changes less than 1\%). Considering that the $\chi^2$/d.o.f\ for some observations was already less than 1 (implying potential overfitting) before adding the reflection component, we still proceed with the \texttt{tbabs*ztbabs*(powerlaw+cflux*thcomp*bbody)} model for our discussion subsequently.}

\begin{deluxetable*}{lccccc}
\tabletypesize{\scriptsize}
\tablewidth{0pt}
\tablecaption{Best-fitting spectral parameters and \(\chi^2/\mathrm{dof}\) values for 9 individual observations and combined soft and hard lag groups}
\tablehead{
  \colhead{Obs.No.} & \colhead{zTBabs} & \colhead{powerlaw} & \colhead{thcomp} &  \colhead{bbody} & \colhead{\(\chi^2/\mathrm{dof}\)} \\
  \colhead{---} & \colhead{\(N_{\mathrm{H},\mathrm{host}}\)(\(\times 10^{20}\,\mathrm{cm^{-2}}\))} & \colhead{$\Gamma$ \& \(F_\mathrm{pl}\)($\times$10$^{-12}$ ergs cm$^{-2}$ s$^{-1}$)} & \colhead{$\Gamma_{\tau}$ \& \(f_{\mathrm{cov}}\)} & \colhead{\(kT_\mathrm{bb}\,(\mathrm{keV})\)
\& \(F_\mathrm{th}\)($\times$10$^{-12}$ ergs cm$^{-2}$ s$^{-1}$)} & \colhead{---}\\
\cline{1-6}
\colhead{(1)} & \colhead{(2)} & \colhead{(3)} & \colhead{(4)} & \colhead{(5)} & \colhead{(6)} 
}
\startdata
Obs2 & $0.7^{+0.3}_{-0.2}$ & $4.56^{+0.14}_{-0.003}$, $6.76^{+0.27}_{-0.29}$ & $2.09^{+0.06}_{-0.02}$, $0.340^{+0.018}_{-0.014}$ & $0.116^{+0.002}_{-0.001}$, $4.32^{+0.32}_{-0.04}$ & 123.16/122 \\
Obs3 & $1.1^{+0.3}_{-0.3}$ & $4.31^{+0.10}_{-0.07}$, $8.27^{+0.30}_{-0.30}$ & $1.87^{+0.08}_{-0.06}$, $0.325^{+0.021}_{-0.016}$ & $0.121^{+0.002}_{-0.001}$, $3.14^{+0.25}_{-0.22}$ & 151.42/119 \\
Obs4 & $1.2^{+0.4}_{-0.3}$ & $4.48^{+0.13}_{-0.10}$, $7.62^{+0.41}_{-0.25}$ & $1.99^{+0.07}_{-0.07}$, $0.371^{+0.019}_{-0.020}$ & $0.124^{+0.002}_{-0.001}$, $3.46^{+0.25}_{-0.25}$ & 92.41/118 \\
Obs5 & $0.7^{+0.2}_{-0.1}$ & $4.64^{+0.17}_{-0.12}$, $6.93^{+0.28}_{-0.19}$ & $2.12^{+0.08}_{-0.07}$, $0.454^{+0.019}_{-0.009}$ & $0.120^{+0.002}_{-0.001}$, $3.37^{+0.33}_{-0.25}$ & 121.05/115 \\
Obs6 & $0.9^{+0.1}_{-0.3}$ & $4.30^{+0.06}_{-0.12}$, $7.99^{+0.21}_{-0.29}$ & $1.82^{+0.05}_{-0.13}$, $0.298^{+0.020}_{-0.045}$ & $0.111^{+0.004}_{-0.004}$, $2.93^{+0.27}_{-0.30}$ & 137.90/118 \\
Obs7 & $1.5^{+0.5}_{-0.5}$ & $4.48^{+0.20}_{-0.15}$, $8.08^{+0.47}_{-0.41}$ & $1.98^{+0.09}_{-0.05}$, $0.339^{+0.028}_{-0.013}$ & $0.114^{+0.003}_{-0.001}$, $3.46^{+0.41}_{-0.27}$ & 124.58/120 \\
Obs8 & $2.0^{+0.1}_{-0.2}$ & $4.72^{+0.06}_{-0.14}$, $8.54^{+0.39}_{-0.35}$ & $2.01^{+0.05}_{-0.11}$, $0.327^{+0.011}_{-0.024}$ & $0.117^{+0.002}_{-0.001}$, $3.89^{+0.12}_{-0.39}$ & 107.55/114 \\
Obs9 & $0.2^{+0.4}_{-0.2}$ & $4.27^{+0.08}_{-0.03}$, $6.96^{+0.49}_{-0.43}$ & $1.73^{+0.10}_{-0.03}$, $0.244^{+0.025}_{-0.014}$ & $0.119^{+0.002}_{-0.001}$, $3.06^{+0.32}_{-0.09}$ & 138.76/113 \\
Obs10 & $0.3^{+0.3}_{-0.2}$ & $4.32^{+0.16}_{-0.12}$, $6.61^{+0.19}_{-0.12}$ & $1.94^{+0.06}_{-0.05}$, $0.294^{+0.032}_{-0.021}$ & $0.120^{+0.003}_{-0.002}$, $3.53^{+0.29}_{-0.20}$ & 163.43/112 \\
Soft & $0.9^{+0.2}_{-0.3}$ & $4.41^{+0.09}_{-0.09}$, $7.47^{+0.23}_{-0.28}$ & $1.96^{+0.05}_{-0.06}$, $0.330^{+0.016}_{-0.020}$ & $0.120^{+0.002}_{-0.002}$, $3.58^{+0.19}_{-0.19}$ & 172.60/122 \\
Hard & $0.8^{+0.3}_{-0.2}$ & $4.46^{+0.11}_{-0.06}$, $7.54^{+0.30}_{-0.23}$ & $1.92^{+0.07}_{-0.04}$, $0.325^{+0.023}_{-0.013}$ & $0.116^{+0.002}_{-0.001}$, $3.27^{+0.48}_{-0.12}$ & 190.87/115 \\
\enddata
\tablecomments{Column definitions are as follows: (1) observation number representing individual observations and combined soft/hard lag groups; (2) self-absorbed equivalent hydrogen column density \(N_{\mathrm{H},\mathrm{host}}\) in units of \(10^{22}\,\mathrm{cm^{-2}}\); (3) power-law photon index $\Gamma$ and power-law flux \(F_\mathrm{pl}\) in units of \(10^{-12}\,\mathrm{ergs\,cm^{-2}\,s^{-1}}\); (4) thermal Comptonization photon power-law index \(\Gamma_{\tau}\) and covering fraction \(f_{\mathrm{cov}}\); (5) blackbody temperature \(kT_\mathrm{bb}\) in\(\,\mathrm{keV}\) and thermal Comptonization flux \(F_\mathrm{th}\) in units of \(10^{-12}\,\mathrm{ergs\,cm^{-2}\,s^{-1}}\); (6) the ratio of chi-squared to degrees of freedom.}
\end{deluxetable*}

\label{table2}

\begin{figure*}[!htbp]
    \centering
    \begin{minipage}{0.8\textwidth}
        \centering
        \includegraphics[width=\textwidth]{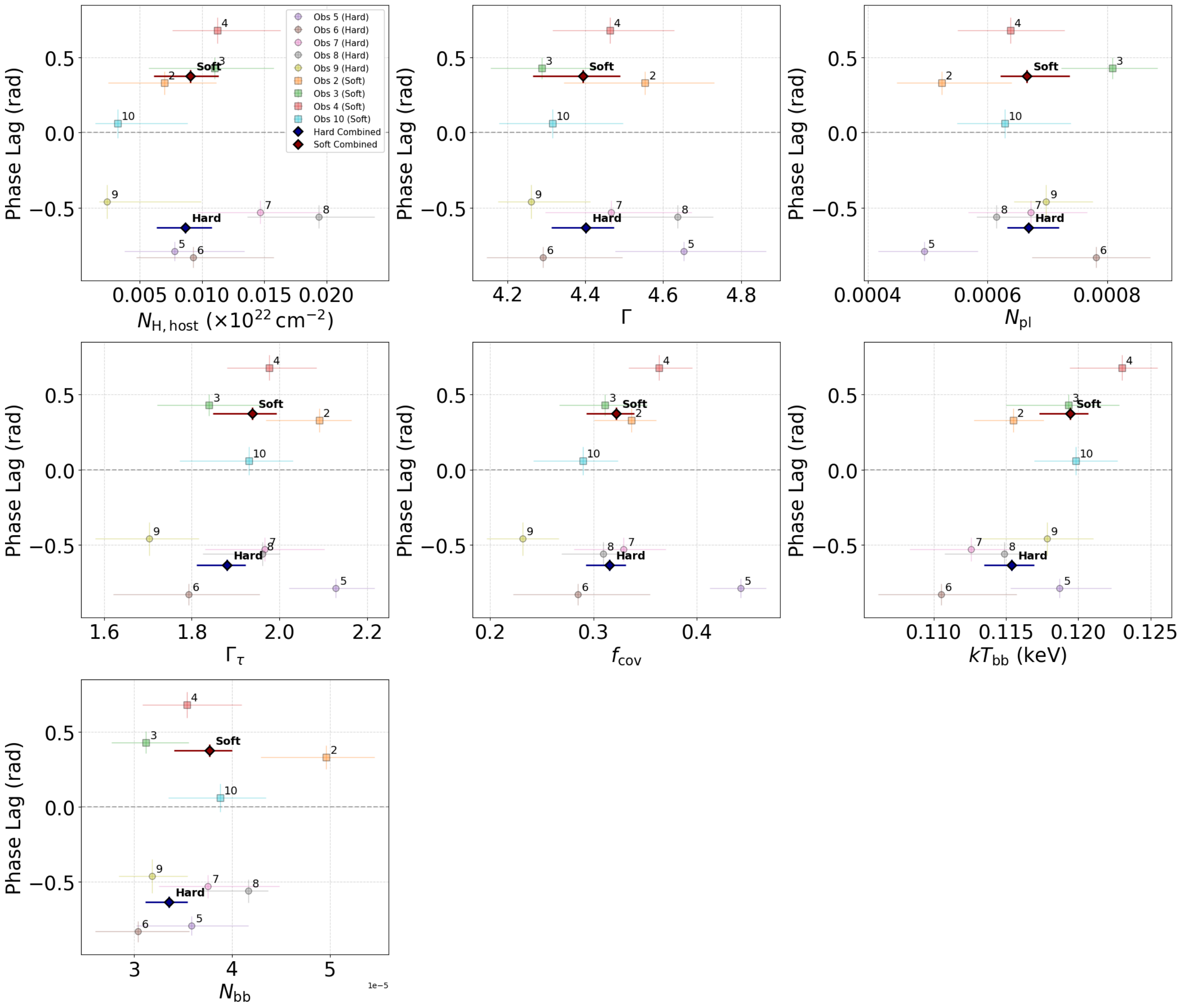}
        \vspace{0.2cm}  
    \end{minipage}
    \hspace{0.1in}

    \caption{%
      Phase lag-spectral parameter relationship plots for self-absorption equivalent hydrogen column density (\(N_{\mathrm{H},\mathrm{host}}\)), photon index ($\Gamma$), power-law flux ($F_{pl}$), Comptonization flux ($F_{th}$), thermal Comptonization photon power-law index ($\Gamma$), covering factor ($f_{cov}$), and blackbody temperature ($kT_{bb}$). Light-colored points represent data points from individual observations, dark points represent data points after group combination. Phase lags are taken as the difference between phase shift values of the \(1\)--\(10\,\mathrm{keV}\) and \(0.3\)--\(1\,\mathrm{keV}\) energy bands, and the errors are calculated using error propagation formulas. Spectral parameter errors represent uncertainties at the \(1\sigma\) confidence level.}
    \label{fig6}
\end{figure*}

Furthermore, we investigated the evolution of spectral parameters among different observations. Figure~\ref{fig6} plots correlations among QPO phase lag and seven free spectral parameters, where the phase lag and its error are the same as in Figure~\ref{fig4}, taken as the difference between phase shift values of the \(1\)--\(10\,\mathrm{keV}\) and \(0.3\)--\(1\,\mathrm{keV}\) energy bands. Each observation corresponds to one data point in each relationship plot, as shown by the light-colored points in Figure~\ref{fig6}. To improve statistics and reduce errors, and examine the differences in spectral parameters under the two lag modes, we combined the average spectra of the soft and hard lag observation groups. We fitted the combined spectra of the soft and hard groups using the same model and fixed parameters, and obtained their spectral parameters and errors using the same method. The phase lag of the soft lag group takes the average value of phase lags from observations \(2\), \(3\), \(4\), and \(10\), while the phase lag of the hard lag group is the average of observations \(5\), \(6\), \(7\), \(8\), and \(9\). The phase lag errors after combination use the square root of the sum of squared errors within the group divided by the sample size. The phase lag-spectral parameter data points after combination are shown as dark-colored points in Figure~\ref{fig6}. It can be seen that the blackbody temperatures in the soft and hard lag modes show marginal difference, with soft lag corresponding to higher blackbody temperature. Other parameters show no significant differences.

\section{Discussion} \label{sec5}
\textbf{Due to observational selection effects, clearly detected QPOs in AGNs are extremely rare.} As the first Active Galactic Nucleus (AGN) with confirmed QPO signals, RE J1034+396 exhibits unique QPO signal strength among AGNs. For this reason, various models have been proposed to explain the generation mechanisms of its QPOs and related phenomena. Building on previous research results, our study mainly found that in the latest batch of observational results from 2020-2021, although the two mutually convertible QPO lag modes (soft lag and hard lag) of RE J1034+396 are consistent in the variation of fractional rms with energy (Figure~\ref{fig3}c), they show differences in hardness ratio and spectral parameters: the soft lag mode corresponds to harder spectra and relatively higher blackbody spectral component temperature (\(kT_\mathrm{bb}\)). Combined with these results, we will discuss several newer models proposed after the discovery that RE~J1034+396 exhibits both soft lag and hard lag, as well as a QPO and lag generation mechanism that we conceived ourselves.

First is the model proposed by \citet{2020MNRAS.495.3538J,2021MNRAS.500.2475J} when QPO phase lag reversal (hard lag) was first observed. They proposed that the QPOs of RE~J1034+396 have intrinsic hard lag, while the previously reported soft lag is due to random variable interference. Through joint spectral-timing analysis, they found that RE~J1034+396 requires four spectral components for complete description: inner disk (\texttt{diskbb}), two warm corona regions at different temperatures (\texttt{CompTT-1} and \texttt{CompTT-2}), and hot corona (\texttt{nthComp}). They proposed that QPOs have layered propagation characteristics: the perturbation signal first originates in the inner disk region, then propagates sequentially to the hotter warm corona region (\texttt{CompTT-2}, delayed by \(679\,\mathrm{s}\) \textbf{(observer frame, approximately \(651\,\mathrm{s}\)  when converted to the AGN rest frame)}), and finally reaches the hot corona region (\texttt{nthComp}, with an additional delay of \(180\,\mathrm{s}\) \textbf{(observer frame, approximately \(173\,\mathrm{s}\)  when converted to the AGN rest frame)}), being progressively amplified during propagation. This inward propagation mechanism explains the observed time lag pattern where soft X-rays lead hard X-rays: soft X-rays mainly originate from the inner disk and hotter warm corona, thus showing leading behavior; while hard X-rays mainly come from the innermost hot corona, receiving propagated perturbation signals, thus appearing lagged. Drawing parallels with the Galactic black hole binary GRS~\(1915+105\), they proposed that QPOs originate from periodic geometric oscillations in the vertical structure of the inner disk edge, resembling the ``plus mode" of torus oscillations. However, this physical model has difficulty matching the latest results from XMM-Newton satellite: it cannot well explain the phenomenon of mutual conversion between soft and hard lag modes observed within relatively short timescales (approximately \(2\) weeks), and it is also inappropriate to attribute all soft lags observed multiple times within half a year to random variable interference.

Second is the model proposed by \citet{2025ApJ...987..135T}. They performed power spectrum and cross-spectrum analysis in the soft (\(0.3\)--\(0.5\,\mathrm{keV}\)) and hard (\(2\)--\(7\,\mathrm{keV}\)) energy bands, and found that the QPO signals observed in the soft energy band actually come from about \(10\%\) contribution of hot corona emission in the low energy range, while the accretion disk itself does not produce QPOs, thus proposing that QPOs completely originate from the hot corona region rather than the traditionally thought accretion disk. The complexity of time lag phenomena stems from two key mechanisms: first is the intrinsic soft lag process---hot corona radiation heats the accretion disk and is reprocessed, producing soft X-ray lag of about \(2000\,\mathrm{s}\) \textbf{(observer frame, approximately \(1920\,\mathrm{s}\)  when converted to the AGN rest frame)}; second is the phase wrapping effect (when the delay time exceeds half a period of the signal period, mathematically it appears as a ``leading" signal rather than a ``lagging" signal)---phase wrapping occurs when QPO frequency reaches approximately \(2.6 \times 10^{-4}\,\mathrm{Hz}\), causing the delay sign to reverse. The observed ``soft lag" and ``hard lag'' phenomena are essentially manifestations of the same physical process at different frequencies. This model is well validated in QPO frequency: among the \(7\) observations in the paper, all those with measured soft lag have QPO frequencies below the critical value, while all those with measured hard lag have QPO frequencies above the critical value. However, our research finds that observations with soft lag have harder spectra (see Figure~\ref{fig4}), while the phase wrapping model predicts similar spectra for the leading and lagging groups. Meanwhile, according to spectral analysis, we see that the soft and hard modes also have a certain degree of correlation with blackbody temperature \(kT_\mathrm{bb}\), which this model has difficulty explaining.

Alternatively, one may also consider another possible picture about disk-corona interaction. The QPOs may originate from an oscillating corona above the black hole and accretion disk \citep[referring to][]{2019MNRAS.489.1957W,2023MNRAS.525..221R}, and have intrinsic hard lag from scattering of X-ray photons by the corona material itself (X-rays are emitted after multiple scatterings within the corona), while soft lag comes from reprocessing of photons emitted from the corona by the disk (hard photons emitted from the corona are reflected by the disk, heating the disk and becoming softened before being directed toward the observer). A corona with a relatively small size hanging up above the disk may have soft lag dominate over the intrinsic hard lag of the corona. As evolution proceeds, the corona material continuously falls (the radial height relative to the black hole and disk plane decreases), and extends toward the disk direction, with its geometric shape gradually becoming flattened. When it descends to a certain degree, the corona will produce some obscuration of the disk, even lying on the disk surface. At this time, the effect of disk reflection on the corona in QPO phase lag becomes weaker, thus revealing the intrinsic hard lag phenomenon. Although this model explains the causes of both soft and hard lag modes, and can constrain the disk-corona distance and the geometric size of the corona itself through  timescales, it has difficulty explaining the hardness ratio phenomenon: when soft X-rays from disk reflection weaken or even completely disappear due to obscuration, there should be a larger hardness ratio, which corresponds to hard lag mode, but this is exactly opposite to our observed result  (Figure~\ref{fig4}).Additionally, based on this model we expect that the covering factor ( \(f_{\mathrm{cov}}\)) should show obvious changes between the soft and hard lag modes, but this is not found in our spectral analysis. Furthermore, this model also has difficulty explaining the correlation between blackbody temperature (\(kT_\mathrm{bb}\)) and lag modes in spectral analysis, as well as the absence of reflection peak components in the spectrum (which can only be explained as strong line broadening).

Additionally, \citet{2025ApJ...989...59X} proposed the relativistic precession model (RPM). They proposed that the QPO emission region of RE~J1034+396 is located on the accretion disk, and a dual-component structure composed of hot corona and warm corona fills in the inner regions of the accretion disk. The entire corona system undergoes periodic precession around the black hole spin axis due to the Lense-Thirring precession effect \citep{1918PhyZ...19..156L} with a period approximately \(92.2\) days \textbf{(observer frame, approximately \(88.5\,\mathrm{days}\)  when converted to the AGN rest frame)}, thus changing the geometric configuration and causing variations in the interaction paths between QPO photons and different coronae. When the corona tilts away from the observer, QPO photons first interact with the warm corona, then with the hot corona, corresponding to a hard lag of about \(300\,\mathrm{s}\) \textbf{(observer frame, approximately \(288\,\mathrm{s}\)  when converted to the AGN rest frame)} with a path difference of about \(30\,R_g\). When the corona is parallel to the disk plane, QPO photons interact less with the warm corona, causing QPO amplitude to decrease significantly with soft X-ray QPOs becoming almost undetectable. At this time, QPO is in a transitional stage between soft and hard lags. When the corona tilts toward the observer, QPO photons first interact with the hot corona, then reach the warm corona, producing a soft lag of over \(400\,\mathrm{s}\) \textbf{(observer frame, approximately \(384\,\mathrm{s}\)  when converted to the AGN rest frame)} with a path difference of about \(40\,R_g\). These three processes repeat continuously with the precession effect, thus generating a periodic cycle of transition between hard lag and soft lag. This model corresponds to the spectral features seen in that paper and the QPO triplet structure obtained from periodic analysis \citep{2025ApJ...989...59X}, and also agrees well with many of our results. For example, for the observation \(1\) in our study (observation ID \(0865010101\)), due to its characteristics in the phase lag-energy relationship plot (Figure~\ref{fig3}b), we consider it may be approximately in the transitional stage between soft lag and hard lag (even though it was grouped into the hard lag group during classification). It shows almost no QPOs in the soft X-ray band while QPOs are significant in hard X-rays, which is consistent with the condition that the corona is parallel to the disk plane given by this model. Additionally, the correlation between hardness ratio and lag modes (soft lag corresponds to harder spectra) could be due to the corona tilting toward the observer during soft lag, with part of the warm corona being obscured by the accretion disk in the line of sight. Since the hot corona is not obscured, the relative proportion of its scattered photons increases, causing spectral hardening. Meanwhile, during soft lag the hot corona tilts toward the observer, but during hard lag the hot corona may be partially obscured by the warm corona in the line of sight, which can also explain the phenomenon that the soft lag group corresponds to relatively higher best-fitting blackbody temperatures in spectral parameters (Figure~\ref{fig6}). Therefore, we suggest that \textbf{this scenario is qualitatively consistent with the observational features} of RE J1034+396, \textbf{although detailed modeling is required to fully verify this hypothesis}.

\section{Conclusions} \label{sec6}

This study analyzed \(10\) XMM-Newton observations of RE~J1034+396 with focus on the coexistence of two modes in QPO phase lags among different energy bands: ``hard lag'' and ``soft lag''. \textbf{We found that, despite significant differences in phase lag characteristics, the fractional root mean square (RMS) in both modes increases with increasing energy in the low-energy region (0.2--2\(\,\mathrm{keV}\)) and shows fluctuations in the hard X-ray band (2--10\(\,\mathrm{keV}\), this may be influenced by the lower flux and signal-to-noise ratio in the high-energy range.} However, the two lag modes are correlated with the spectral hardness: the soft lag mode corresponds to harder spectra, while the hard lag mode corresponds to softer spectra. Additionally, spectral fitting results show that the blackbody spectral component temperature (\(kT_\mathrm{bb}\)) is relatively higher in the soft lag mode. Based on these results, \textbf{\textbf{our qualitative comparison of different models suggests that}} the relativistic precession model (RPM) \textbf{\textbf{is a promising scenario to interpret the complex timing and spectral behaviors}}.

\begin{acknowledgments}
\textbf{We thank the anonymous referee for valuable comments that made the paper more complete and rigorous.} And we thank the High Energy Astrophysics Science Archive Research Center (HEASARC) at NASA's Goddard Space Flight Center for providing the data. This work is supported by the National Key R\&D Program of China (2021YFA0718500) and the National Natural Science Foundation of China (NSFC) under Grants No. 12025301 and 12333007. It is also partially supported by the International Partnership Program of the Chinese Academy of Sciences (Grant No. 113111KYSB20190020). We acknowledeg support from China's Space Origins Exploration Program (eXTP).
\end{acknowledgments}

\appendix  

\section{MCMC Parameter Probability Distributions}\label{appendix a}

\begin{figure*}[!htbp]
    \centering
    \begin{minipage}{0.8\textwidth}
        \centering
        \includegraphics[width=\textwidth]{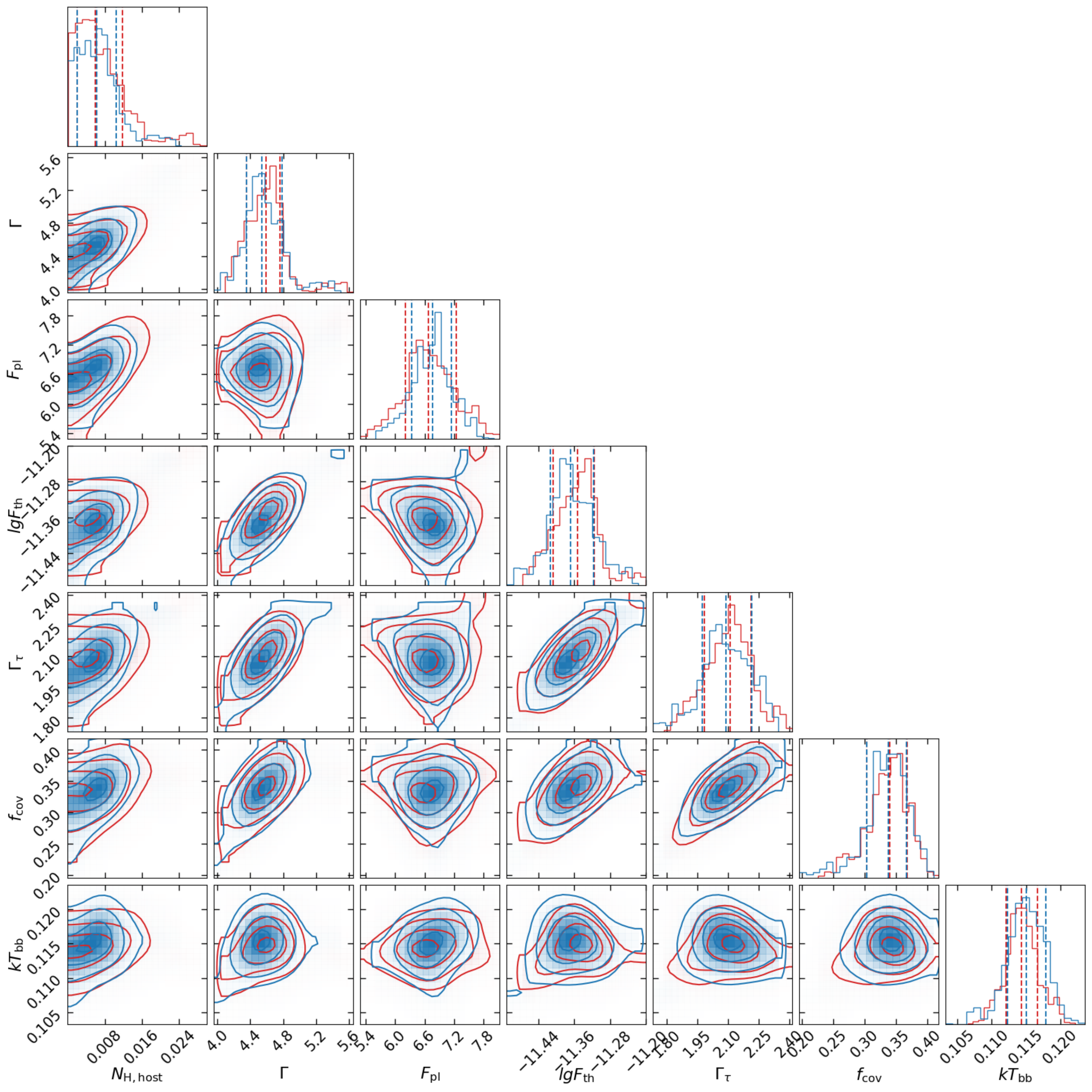}
        \vspace{0.2cm}  
    \end{minipage}
    \hspace{0.1in}

    \caption{%
      One- and two-dimensional projections of the posterior probability distributions for observation \(2\), and the 0.16, 0.5, and 0.84 quantile contours derived from the MCMC analysis for each free spectral parameter (\(N_{\mathrm{H},\mathrm{host}}\)(\(\times 10^{22}\,\mathrm{cm^{-2}}\)),$\Gamma$, \(F_\mathrm{pl}\)($\times$10$^{-12}$ ergs cm$^{-2}$ s$^{-1}$), \(lgF_\mathrm{th}\)($\times$10$^{-12}$ ergs cm$^{-2}$ s$^{-1}$), $\Gamma_{\tau}$, \(f_{\mathrm{cov}}\), \(kT_\mathrm{bb}\,(\mathrm{keV})\)), from the joint spectral fitting of data from the European Photon Imaging Cameras (EPIC). To test the convergence, we compare the one- and two-dimensional projections of the posterior distributions from the first (red) and second (blue) halves of the chain. This illustration corresponds to the spectral fitting of the QPO peak phase.}
    \label{fig7}
\end{figure*}

This appendix contains corner plots of spectral parameters from an example MCMC analysis for observation \(2\). We use the Goodman-Weare algorithm with \(20\) walkers and a total length of \(10000\) to perform the MCMC analysis, and the initial \(2000\) elements are discarded as the burn-in period during which the chain reaches its stationary state. In Figure~\ref{fig7} we compare the one- and two-dimensional projections of the posterior distributions for each parameter from the first and second halves of the chain to test the convergence. The contour maps and probability distributions are plotted using the corner package \citep{2016JOSS....1...24F}. 

%




\bibliography{paper}{}
\bibliographystyle{aasjournal}



\end{document}